\documentclass{ws-procs9x6}

\begin{document}

\title{Design Considerations and Sensitivity Estimates for an Acoustic
  Neutrino Detector\footnote{\uppercase{T}his work was supported by
    the \uppercase{G}erman \uppercase{BMBF} \uppercase{G}rant
    \uppercase{N}o. 05 \uppercase{CN2WE1/2}.}}

\author{T. KARG, G. ANTON, K. GRAF, J. H\"OSSL, A.  KAPPES, U. KATZ, \\
  R. LAHMANN, C. NAUMANN and K. SALOMON}

\address{Physikalisches Institut, \\
  Friedrich-Alexander-Universit\"at Erlangen-N\"urnberg, \\
  Erwin-Rommel-Stra\ss e 1, \\
  91058 Erlangen, Germany \\
  E-mail: Timo.Karg@physik.uni-erlangen.de}

\maketitle

\abstracts{We present a Monte Carlo study of an underwater neutrino
  telescope based on the detection of acoustic signals generated by
  neutrino induced cascades. This provides a promising approach to
  instrument large detector volumes needed to detect the small flux of
  cosmic neutrinos at ultra-high energies ($E \gtrsim 1 \,
  \mathrm{EeV}$). Acoustic signals are calculated based on the
  thermo-acoustic model.  The signal is propagated to the sensors
  taking frequency dependent attenuation into account, and detected
  using a threshold trigger, where acoustic background is included as
  an effective detection threshold. A simple reconstruction algorithm
  allows for the determination of the cascade direction and energy.
  Various detector setups are compared regarding their effective
  volumes.  Sensitivity estimates for the diffuse neutrino flux are
  presented.}

\section{Introduction}

Very large target masses are required to detect the low neutrino
fluxes predicted at ultra-high energies. Current water \v{C}erenkov
neutrino telescopes (AMANDA, BAIKAL, ANTARES, NESTOR, ...) and
next-generation $\mathrm{km}^3$ size detectors (IceCube, KM3NeT) do
not have sufficient fiducial volume to detect, for example, GZK
neutrinos. The affordable size of these detectors is limited by the
attenuation length of light in water or ice which restricts the
spacing between optical sensors.
G.A. Askariyan described a hydrodynamic mechanism of sound generation
for charged particles propagating through water\cite{Askariyan} which
can be exploited for an acoustic neutrino telescope. The
thermo-acoustic model has since been verified in the laboratory
several times and with high
precision\cite{Sulak}$^,$\cite{Hunter}$^,$\cite{Erlangen}.  Utilizing
the fact that, for the frequencies considered, the sonic attenuation
length in water is about ten times larger than the optical attenuation
length, much larger volumes could be instrumented with the same number
of sensors.
In the next section we describe the simulation chain used for studying
acoustic neutrino telescopes. After that, sensitivity estimates for an
acoustic detector are derived.

\section{The simulation chain}

For the simulation an isotropic flux of highest-energy neutrinos
($10^8 \, \mathrm{GeV} < E_\nu < 10^{16} \, \mathrm{GeV}$) is
generated. Equal numbers of neutrinos are produced in each energy bin
of constant width in $\log E$, with a given energy spectrum following
a power law ($E^{-2}$) in each $E$ bin. It is assumed that all
neutrinos from above can propagate freely down to the detector. On the
other hand, the earth is assumed to be opaque for all neutrinos coming
from below the horizon.  The elasticity distribution of the neutrino
interaction is taken from the ANIS neutrino interaction
simulator\cite{Anis}. For electromagnetic cascades the LPM effect,
which elongates the cascade and thus reduces the energy density and
the amplitude of the acoustic signal, has to be taken into account.
Since there is no reliable shower simulation including the LPM effect
in water so far, the leptonic branch of all neutrino interactions is
discarded, even for electron-neutrino charged-current interactions.

The three-dimensional cascade development and energy deposition were
studied with GEANT4 up to primary hadronic energies of $100 \,
\mathrm{TeV}$ using the QGSP interaction model. The shape and the
spatial extension of the energy distribution were found to vary only
slightly with the primary energy. Therefore, the spatial distribution
of the energy is assumed to be the same for all energies, and the
energy density scales linearly with the energy of the hadronic system.
This energy distribution and the thermodynamic parameters of water are
then used as an input to the thermo-acoustic model which gives the
resulting bipolar acoustic signal for every sensor position. The
amplitude of the bipolar pulse depends on the cascade energy only.

\begin{figure}[ht]
  \centerline{\epsfxsize=68mm\epsfbox{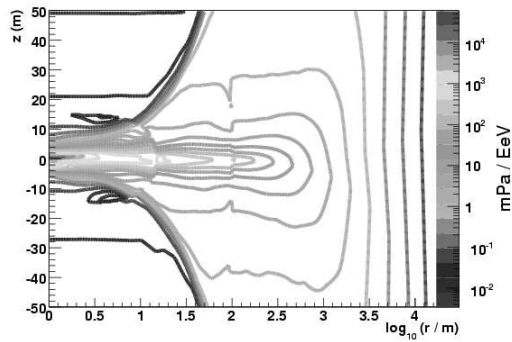}}
  \caption{Parameterization of the amplitude of the sonic field for a
    hadronic cascade centered at the origin. The cascade has a length
    of approx. $15 \, \mathrm{m}$ and develops in positive $z$
    direction. \label{fig1}}
\end{figure}

Sonic attenuation in sea water is strongly frequency dependent. The
attenuation length for the typical signal frequency of approx. $20 \,
\mathrm{kHz}$ is $1 \, \mathrm{km}$ (compared to $50$ -- $70 \,
\mathrm{m}$ optical attenuation length relevant for water \v{C}erenkov
neutrino telescopes).  It is accounted for by applying a frequency
filter to the acoustic signal at a given sensor position.
Figure~\ref{fig1} shows the parameterization of the amplitude of the
bipolar signal as a function of position, which is used in the
simulation to determine the sensor response for a given hadronic
cascade.

The smallest unit of the simulated acoustic detector is an ``acoustic
module'' (AM) which is a device that can detect bipolar acoustic
signals above a given detection threshold, $p_\mathrm{th}$, determined
by the background noise in the sea. Such an AM might be realized as a
local array of hydrophones allowing the suppression of background with
short correlation length. According to Ref.~\refcite{Danaher} a
threshold of $35 \, \mathrm{mPa}$ has to be used for a single
hydrophone if one allows for one false signal in $10$ years at a
five-fold coincidence. Using AMs consisting of multiple hydrophones
should allow to lower this threshold down to $5 \, \mathrm{mPa}$.

Our detector consists of AMs that are arranged randomly inside the
instrumented volume in order to avoid geometrical effects on the
sensitivity estimates. Neutrino events are generated homogeneously and
with $2 \pi \, \mathrm{sr}$ angular distribution in a volume with a
height of $2.5 \, \mathrm{km}$ (corresponding to typical depths in the
Mediterranean Sea), and a radius of $10 \, \mathrm{km}$; the resulting
generation volume is denoted by $V_\mathrm{gen}$. Each AM records the
arrival time and amplitude of the signal if it is above the threshold
$p_\mathrm{th}$.  An event is triggered if four or more AMs detect a
signal. For our study a timing resolution of $10 \, \mathrm{\mu s}$
($100 \, \mathrm{kHz}$ sampling frequency), a positioning accuracy of
$10 \, \mathrm{cm}$ for the AMs and an amplitude resolution of $2 \,
\mathrm{mPa}$ are implied, which are all realized by Gaussian
smearing.

The shower reconstruction is performed in two steps. First, the shower
position is reconstructed by minimization of the residuals of the
arrival times assuming an isotropic sonic point source (which is a
valid assumption since the typical inter-AM distance is large compared
to the shower extension). With this method the cascades center of
gravity can be reconstructed with a RMS of $14 \, \mathrm{cm}$ in each
Cartesian coordinate. Based on this position and the parameterization
of the sonic field (Fig.~\ref{fig1}) the direction and energy of the
cascade are reconstructed by minimizing the amplitude residuals.
Without applying any selection cuts the median of the error in the
direction reconstruction is $7^\circ$, where events are still
included, for which the reconstruction seems to fail completely. The
energy can be determined up to a factor of $3$.

\section{Sensitivity estimates}

\begin{figure}[ht]
  \centerline{\epsfxsize=64mm\epsfbox{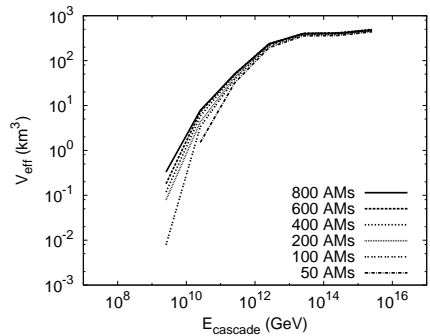}}
  \caption{Effective volume for various sensor densities
    (instrumented volume: $1 \, \mathrm{km}^3$). \label{fig2}}
\end{figure}

Based on the detector simulation chain presented above it is possible
to derive sensitivity estimates for various detector configurations.
We use the effective volume defined as $V_\mathrm{eff} =
\frac{N_\mathrm{reco}}{N_\mathrm{gen}} \, V_\mathrm{gen}$ as a measure
for the sensitivity of a detector, where $N_\mathrm{reco}$ is the
number of reconstructed events (reconstruction fits converge) without
any selection cuts obtained from $N_\mathrm{gen}$ events generated
inside the volume $V_\mathrm{gen}$. Figure~\ref{fig2} shows the effect
of varying the instrumentation density of the detector between $50$
and $800 \, \mathrm{AM} / \mathrm{km}^3$. For densities much lower
than approx. $200 \, \mathrm{AM} / \mathrm{km}^3$ the effective volume
drops dramatically at lower energies, and thus, the lower energy
threshold rises.

\begin{figure}[ht]
  \centerline{\epsfxsize=64mm\epsfbox{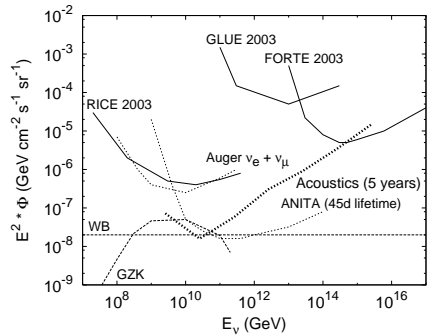}}
  \caption{Neutrino flux limit derived for a $30 \times 50
    \times 1 \, \mathrm{km}^3$ detector with a lifetime of $5$ years.
    Dashed lines are theoretical models (extrapolated Waxman-Bahcall
    flux and GZK neutrinos). Solid lines are
    experimental flux limits; dotted lines are expected flux limits
    from future experiments. \label{fig3}}
\end{figure}

Further, it is essential for a future acoustic detector to have a
pressure threshold $p _\mathrm{th}$ as low as possible, where the
lower limit is given by the intrinsic background noise in the sea
which is approx. $1 \, \mathrm{mPa}$ (sea state $0$). On the other
hand, a density of only $200 \, \mathrm{AM} / \mathrm{km}^3$ seems
sufficient which allows to instrument very large volumes with a
moderate number of DAQ channels read out at low frequencies ($100 \,
\mathrm{kHz}$), leading to manageable data rates.
In figure~\ref{fig3} we show that, with a detector with $3 \cdot 10^5$
DAQ channels ($30 \times 50 \times 1 \, \mathrm{km}^3$, $200 \,
\mathrm{AM} / \mathrm{km}^3$, $p_\mathrm{th} = 5 \, \mathrm{mPa}$),
several theoretical models that predict neutrinos above $1 \,
\mathrm{EeV}$ could be verified within $5$ years of runtime.

\section{Conclusions}

Acoustic detection is a promising approach to detect cosmic neutrinos
at highest energies. Detectors build of ``acoustic modules'' that can
detect bipolar acoustic signals above $5 \, \mathrm{mPa}$ are able to
reconstruct neutrino-events with energies above $1 \, \mathrm{EeV}$
with as few as $200 \, \mathrm{AM} / \mathrm{km}^3$. This allows for
the construction of a teraton detector, which is necessary to detect
the small neutrino fluxes predicted by theoretical models within a
reasonable time.


\begin{thebibliography}{0}


\bibitem{Askariyan}
G.~A. Askariyan, {\it Atomnaya Energiya} {\bf 3}, 152 (1957). \\
G.~A. Askariyan et al., {\it Nucl. Inst. Meth.} {\bf 164}, 267 (1979).

\bibitem{Sulak}
L. Sulak et al., {\it Nucl. Inst. Meth.} {\bf 161}, 203 (1979).

\bibitem{Hunter}
S.~D. Hunter et al., {\it J. Acoust. Soc. Am.} {\bf 69}, 1557 (1981).

\bibitem{Erlangen}
K. Graf et al., 1st International ARENA Workshop, Zeuthen (2005). 

\bibitem{Anis}
M. Kowalski and A. Gazizov, 28th ICRC, Tsukuba, 1459 (2003).

\bibitem{Danaher}
S. Danaher et al., 1st International ARENA Workshop, Zeuthen (2005).

\end{thebibliography}
\end{document}